\documentclass{emulateapj}
\usepackage[]{graphicx}

\shorttitle{Virgo Galaxies with Long One-Sided H{\small I} Tails}
\shortauthors{Chung et al.}
\begin{document}

\newcommand{\hi}{\mbox{H{\sc i}}}
\newcommand{\shi}{$S_{\tiny\mbox{H{\sc i}}}$}
\newcommand{\mhi}{$M_{\tiny\mbox{H{\sc i}}}$}

\title{Virgo Galaxies with Long One-Sided H{\small I} Tails}

\author{Aeree Chung\altaffilmark{1}, J. H. van Gorkom\altaffilmark{2},
         Jeffrey D. P. Kenney\altaffilmark{3} and Bernd Vollmer\altaffilmark{4}}
\altaffiltext{1}{Department of Astronomy, Columbia University, 550 West 120th Street, New York,
                 NY~10027, U.S.A.; now NRAO Jansky Fellow at UMass, achung@astro.umass.edu}
\altaffiltext{2}{Department of Astronomy, Columbia University, 550 West 120th Street, New York,
                 NY~10027, U.S.A.; jvangork@astro.columbia.edu}
\altaffiltext{3}{Department of Astronomy, Yale University, P.O. Box 208101, New Haven, 
                 CT 06520, U.S.A.; kenney@astro.yale.edu}
\altaffiltext{4}{Observatoire astronomique de Stasbourg, 11 rue de 
                 l'Universite, F-67000 Strasbourg, France; 
                 bvollmer@astro.u-strasbg.fr}

\begin{abstract}
In a new H{\sc i} imaging survey of Virgo galaxies (VIVA: VLA Imaging of Virgo galaxies in 
Atomic gas), we find 7 spiral galaxies with long H{\sc i} tails. The morphology varies but 
all the tails are extended well beyond the optical radii on one side. These galaxies are found 
in intermediate-low density regions ($0.6-1$~Mpc in projection from M87). The tails are all
pointing roughly away from M87, suggesting that these tails may have been created by a global
cluster mechanism. While the tidal effects of the cluster potential are too small, a rough 
estimate suggests that simple ram-pressure stripping indeed could have formed the tails in all 
but two cases. At least three systems show  H{\sc i} truncation to within the stellar disk, 
providing evidence for a gas-gas interaction. Although most of these galaxies do not appear 
disturbed optically, some have close neighbors, suggesting that tidal interactions 
may have moved gas outwards making it more susceptible to the ICM ram-pressure or viscosity. 
Indeed, a simulation study of one of the tail galaxies, NGC 4654, suggests that the galaxy is
most likely affected by the combined effect of a gravitational interaction and ram-pressure 
stripping. We conclude that these one-sided H{\sc i} tail galaxies have recently arrived in 
the cluster, falling in on highly radial orbits. It appears that galaxies begin to lose their
gas already at intermediate distances from the cluster center through ram-pressure or 
turbulent viscous stripping and tidal interactions with neighbours, or a combination of both.
\end{abstract}
 
\keywords{galaxies: clusters --- galaxies: evolution --- galaxies: interactions 
--- galaxies: kinematics and dynamics}

\section{Introduction}
\label{sec-intro}
The density-morphology relation \citep{dressler80}, indicating an increasing elliptical and
S0 population fraction with increasing density, has been known to exist over six orders of 
magnitude in galaxy density \citep{pg84}. It is only recently that large surveys have shown 
just how smoothly the galaxy properties change with galaxy density. For example 
\citet{solanes01} find that gas deficiency gradually decreases out to two Abell radii from 
the cluster center and \citet{lewis02} and \citet{gomez03} find that the star formation rate 
depends on local galaxy density. These results have led to the concept that galaxies may 
already be affected by their environment well before they fall into the dense cluster core
({\it pre-processing}).

While many processes have been suggested to affect galaxies in the cluster environment such 
as ram-pressure stripping \citep{gg72}, starvation \citep{ltc80}, harassment \citep{moo96} 
and tidal distortions due to the cluster potential and galaxy-galaxy interactions 
\citep{bekki99,mihos04}, it is still an open question what would affect the galaxies in the
lower density outskirts of the clusters. Recently \citet{kenney04} studied one such case,
NGC~4522, which is at a large projected distance from the center of Virgo, yet appears to 
be undergoing ram-pressure stripping due to bulk motions of the ICM, as the M49 subcluster
is merging with Virgo proper.

In our new VLA H{\sc i} survey of the Virgo cluster, we find candidates that appear to be
feeling the cluster influence for the first time in the outskirts of the cluster. A number 
of galaxies located at intermediate distances from M87 reveal long gas tails, all pointing
roughly away from the cluster center. Here we test our assumption that these tail galaxies
are recent arrivals in the cluster and review possible causes of their gas tails, i.e. how
galaxies can be modified before they enter the dense cluster core.

\section{VIVA, a New VLA H{\sc i} Survey of Virgo Spirals}
\label{sec-viva}

VIVA, VLA\footnotemark Imaging of Virgo galaxies in Atomic gas, is a new VLA H{\sc i} 
survey of spiral galaxies in the Virgo cluster. The goal of the survey is to study details
of different environmental effects. The Virgo cluster is ideal for this study not only due
to its nearness but also because it is dynamically young \citep{binggeli99} and very likely
to contain various mechanisms at work. We have sampled $\sim$50 galaxies throughout the
cluster from near the dense core to the outskirts. The selected galaxies are located at
projected distances of 0.3-3.3 Mpc from the cluster center assuming a distance of 16~Mpc
to Virgo \citep[][]{yasuda97}, which corresponds to $\approx0.4$-4.3 virial radii 
\citep{ts84}.

\footnotetext{The VLA is operated by the National Radio Astronomy Observatory, which is a 
facility of the National Science Foundation (NSF), operated under cooperative agreement 
by Associated Universities, Inc.} 

The new survey was done with the VLA in the C-short array and additional 10 fields (13 
galaxies) previously observed in C array were also included in the database. The typical 
spatial and spectral resolutions are 15$''$ and 10 km~s$^{-1}$, respectively, which are a
factor of 3 and 4 better than the previous VLA H{\sc i} survey of Virgo spirals 
\citep{cayatte90}. The rms of the data is typically 0.4 mJy, yielding an H{\sc i}
column density sensitivity of $\approx6\times10^{19}$ cm$^{-2}$ in 3$\sigma$.

The data have been calibrated and continuum subtracted with the Astronomical Imaging
Processing System (AIPS). The data from the NRAO archive have been processed in the same
way as the new survey data. To maximize the sensitivity, we applied a weighting scheme
intermediate between uniform and natural but somewhat closer to a natural weighting scheme
\citep[robust=1,][]{briggs95}. We also made tapered images with twice the beamsize
to bring out faint extended features. We will present further details of the observations
and the data reduction in Chung et al. (2007, in preparation).

%\notetoeditor{Table should appear on the top of pg. 3}

\begin{table*}
\begin{center}
\label{tbl-tail}
\scriptsize
\caption{\sc Properties of the Virgo Hi Tail Galaxies}
\begin{tabular}{llrrrrrcrrrrrrrrlrrc}
\hline\hline
\noalign{\vspace{0.1cm}} &
\multicolumn{4}{c}{\hspace{-0.2cm}-- General properties --} &
\multicolumn{8}{c}{-------------------------- H{\sc i} properties --------------------------} &
\multicolumn{3}{c}{-- Global parameters$^{\rm b}$ --} &
\multicolumn{4}{c}{\hspace{-0.1cm}------ Neighbor ------}\\

\multicolumn{1}{l}{ } &  
\multicolumn{1}{l}{\hspace{0cm}Type} & 
\multicolumn{1}{c}{\hspace{0cm}$B_T$} & 
\multicolumn{1}{c}{\hspace{0cm}$V_{rad}$} &
\multicolumn{1}{c}{\hspace{0cm}$D_{25}$} & 
\multicolumn{1}{c}{\hspace{0cm}$D_{_{\rm HI}}$} &
\multicolumn{1}{c}{\hspace{0cm}$M_{_{\rm HI}}$} &
\multicolumn{1}{c}{\hspace{0cm}$V_{rot}$$^{\rm a}$} &
\multicolumn{1}{c}{\hspace{0cm}$def_{_{\rm HI}}$} &
\multicolumn{1}{c}{\hspace{0cm}$l_{_{tail}}$} & 
\multicolumn{1}{c}{\hspace{0cm}$w_{_{tail}}$} & 
\multicolumn{1}{c}{\hspace{0cm}$M_{_{tail}}$} &
\multicolumn{1}{c}{\hspace{0cm}$\Sigma_{_{tail}}$} &  
\multicolumn{1}{c}{\hspace{0cm}$d_{_{87}}$} &
\multicolumn{1}{c}{\hspace{0cm}$\rho_{_{\rm ICM}}$} &
\multicolumn{1}{c}{\hspace{0cm}$M$($d_{_{87}}$)} &
\multicolumn{1}{c}{\hspace{0cm}} &
\multicolumn{1}{c}{\hspace{0cm}$\Delta v$} &
\multicolumn{1}{c}{\hspace{0cm}$\Delta d$} &
\multicolumn{1}{c}{\hspace{0cm}$\Delta B_T$} \\

\multicolumn{1}{l}{NGC} & 
\multicolumn{1}{l}{\hspace{0cm}} & 
\multicolumn{1}{c}{\hspace{0cm}mag} & 
\multicolumn{1}{c}{\hspace{0cm}km\hspace{-0.05cm}/\hspace{-0.03cm}s} &
\multicolumn{1}{c}{\hspace{0cm}kpc} & 
\multicolumn{1}{c}{\hspace{0cm}kpc} &
\multicolumn{1}{c}{\hspace{0cm}${10^8}$\hspace{-0.05cm}$M_\odot$} &
\multicolumn{1}{c}{\hspace{0cm}km\hspace{-0.05cm}/\hspace{-0.03cm}s} &
\multicolumn{1}{c}{\hspace{0cm}} &
\multicolumn{1}{c}{\hspace{0cm}kpc}& 
\multicolumn{1}{c}{\hspace{0cm}kpc}& 
\multicolumn{1}{c}{\hspace{0cm}${10^8}$\hspace{-0.05cm}$M_\odot$} &
\multicolumn{1}{c}{\hspace{0cm}$10^{19}$\hspace{-0.1cm}/\hspace{-0.05cm}cm$^{2}$} &
\multicolumn{1}{c}{\hspace{0cm}Mpc} &
\multicolumn{1}{c}{\hspace{0cm}$10^{-4}$\hspace{-0.1cm}/\hspace{-0.05cm}cm$^{3}$} &
\multicolumn{1}{c}{\hspace{0cm}$10^{14}\hspace{-0.05cm}M_\odot$} &
\multicolumn{1}{l}{\hspace{0cm}{\tiny}} &
\multicolumn{1}{c}{\hspace{0cm}km\hspace{-0.05cm}/\hspace{-0.03cm}s} &
\multicolumn{1}{c}{\hspace{0cm}kpc} &
\multicolumn{1}{c}{\hspace{0cm}{mag}}\\
\noalign{\vspace{0.1cm}}
\hline
\noalign{\vspace{0.1cm}}
4294&
\hspace{0cm}SBc &       %type
\hspace{0cm}12.3 &      %Bt 
\hspace{0cm}359~&       %Vrad
\hspace{0cm}14~~&       %D25
\hspace{0cm}17~ &       %DHI
\hspace{0cm}15.9~~~~&   %MHI
\hspace{0cm}172 &       %Vrot
\hspace{0cm}$-0.17$~~~ &%DefHI
\hspace{0cm}27~~&       %lTail
\hspace{0cm}8.2 &       %wTail 
\hspace{0cm}2.0~~~~&    %MTail
\hspace{0cm}11-26~~~&   %SDTail
\hspace{0cm}0.7~~&      %d87
\hspace{0cm}1.04~~~~~&  %dICM
\hspace{0cm}0.97~~~~ &  %Mcl
\hspace{0cm}~~N4299&    %nearby
\hspace{0cm}127&        %delV
\hspace{0cm}27&         %delD
\hspace{0cm}0.2\\       %Del magOB

4299&
\hspace{0cm}SABd &      %type
\hspace{0cm}12.5 &      %Bt 
\hspace{0cm}237~&       %Vrad
\hspace{0cm} 7~~&       %D25
\hspace{0cm}12~ &       %DHI
\hspace{0cm}11.8~~~~&   %MHI
\hspace{0cm}169 &       %Vrot
\hspace{0cm}$-0.56$~~~ &%DefHI
\hspace{0cm}14~~&       %lTail
\hspace{0cm}9.0 &       %wTail 
\hspace{0cm}2.6~~~~&    %MTail
\hspace{0cm}11-18~~~&   %SDTail
\hspace{0cm}0.7~~&      %d87
\hspace{0cm}1.09~~~~~&  %dICM
\hspace{0cm}0.93~~~~ &  %Mcl
\hspace{0cm}~~N4294&    %nearby
\hspace{0cm}127&        %delV
\hspace{0cm}27&         %delD
\hspace{0cm}0.2\\       %Del mag

4302&
\hspace{0cm}Sc &        %type
\hspace{0cm}12.5 &      %Bt 
\hspace{0cm}1150~&      %Vrad
\hspace{0cm}24~~&       %D25
\hspace{0cm}26~ &       %DHI
\hspace{0cm}14.9~~~~&   %MHI
\hspace{0cm}199 &       %Vrot
\hspace{0cm}$ 0.39$~~~ &%DefHI
\hspace{0cm}16~~&       %lTail
\hspace{0cm}8.9 &       %wTail 
\hspace{0cm}0.9~~~~&    %MTail
\hspace{0cm}10-20~~~&   %SDTail
\hspace{0cm}0.9~~&      %d87
\hspace{0cm}0.74~~~~~&  %dICM
\hspace{0cm}1.24~~~~ &  %Mcl
\hspace{0cm}~~N4298&    %nearby
\hspace{0cm}14&         %delV
\hspace{0cm}11&         %delD
\hspace{0cm}0.5\\       %Del mag

4330&
\hspace{0cm}Sc &        %type
\hspace{0cm}13.1 &      %Bt 
\hspace{0cm}1564~&      %Vrad
\hspace{0cm}25~~&       %D25
\hspace{0cm}13~ &       %DHI
\hspace{0cm}4.1~~~~&    %MHI
\hspace{0cm}180 &       %Vrot
\hspace{0cm}$ 0.80$~~~ &%DefHI
\hspace{0cm}13~~&       %lTail
\hspace{0cm}5.0 &       %wTail 
\hspace{0cm}0.4~~~~&    %MTail
\hspace{0cm} 6-13~~~&   %SDTail
\hspace{0cm}0.6~~&      %d87
\hspace{0cm}1.19~~~~~&  %dICM
\hspace{0cm}0.86~~~~ &  %Mcl
\hspace{0cm}~~V0706&    %nearby
\hspace{0cm}75&         %delV
\hspace{0cm}77&         %delD
\hspace{0cm}4.3\\       %Del mag

4396&
\hspace{0cm}Scd &       %type
\hspace{0cm}13.1 &      %Bt 
\hspace{0cm}-125~&      %Vrad
\hspace{0cm}15~~&       %D25
\hspace{0cm}14~ &       %DHI
\hspace{0cm}9.9~~~~&    %MHI
\hspace{0cm}170 &       %Vrot
\hspace{0cm}$ 0.28$~~~ &%DefHI
\hspace{0cm}13~~&       %lTail
\hspace{0cm}3.6 &       %wTail 
\hspace{0cm}0.7~~~~&    %MTail
\hspace{0cm}19-39~~~&   %SDTail
\hspace{0cm}1.0~~&      %d87
\hspace{0cm}0.64~~~~~&  %dICM
\hspace{0cm}1.37~~~~ &  %Mcl
\hspace{0cm}~~N4419&    %nearby
\hspace{0cm}133&        %delV
\hspace{0cm}186&        %delD
\hspace{0cm}1.1\\       %Del mag

4424&
\hspace{0cm}SBa &       %type
\hspace{0cm}12.4 &      %Bt 
\hspace{0cm}439~&       %Vrad
\hspace{0cm}16~~&       %D25
\hspace{0cm}10~ &       %DHI
\hspace{0cm}1.7~~~~&    %MHI
\hspace{0cm}158 &       %Vrot
\hspace{0cm}$ 0.75$~~~ &%DefHI
\hspace{0cm}22~~&       %lTail
\hspace{0cm}10.7 &      %wTail 
\hspace{0cm}0.8~~~~&    %MTail
\hspace{0cm}3-6~~~&     %SDTail
\hspace{0cm}0.9~~&      %d87
\hspace{0cm}0.77~~~~~&  %dICM
\hspace{0cm}1.21~~~~ &  %Mcl
\hspace{0cm}~~N4445&    %nearby
\hspace{0cm}79&         %delV
\hspace{0cm}74&         %delD
\hspace{0cm}1.2\\       %Del mag

4654&
\hspace{0cm}Sc(R) &     %type
\hspace{0cm}11.0 &      %Bt 
\hspace{0cm}1038~&      %Vrad
\hspace{0cm}23~~&       %D25
\hspace{0cm}28~ &       %DHI
\hspace{0cm}34.3~~~~&   %MHI
\hspace{0cm}300 &       %Vrot
\hspace{0cm}$0.06$~~~ & %DefHI
\hspace{0cm}32~~&       %lTail
\hspace{0cm}12.9 &      %wTail 
\hspace{0cm}4.1~~~~&    %MTail
\hspace{0cm}20-47~~~&   %SDTail
\hspace{0cm}0.9~~&      %d87
\hspace{0cm}0.68~~~~~&  %dICM
\hspace{0cm}1.32~~~~ &  %Mcl
\hspace{0cm}~~N4639&    %nearby
\hspace{0cm}27&         %delV
\hspace{0cm}81&         %delD
\hspace{0cm}1.2\\       %Del mag
\hline
\noalign{\vspace{0.1cm}}
\multicolumn{20}{l}{$^{\rm a}$It was determined from H{\sc i} position-velocity cut along 
the major axis except for NGC~4424, which was estimated from $H$-band mag.}\\
\multicolumn{20}{l}{$^{\rm b}$$\rho_{\rm ICM}$ and $M(d_{87})$ have been estimated assuming 
the $\beta$-model \citep{schindler99} with the same coefficients for the} \\
\multicolumn{20}{l}{~~Virgo cluster presented as in \citet{vollmer01}.}\\
\end{tabular}
\end{center}
\end{table*}

\section{H{\sc I} Tail Galaxies in the Virgo Cluster}
\label{sec-tails}
\subsection{Overview of the H{\small I} Properties of the Tails}
The survey collected H{\sc i} data on 53 galaxies in total, 48 spirals and 5 systems of other 
types (Im, dE). Among them, 7 spiral galaxies revealed a long H{\sc i} 
extension (see Plate~1). These tail galaxies have the following properties in common in the 
H{\sc i} morphology: 1) the H{\sc i} is extended well beyond the optical disk only on one side;
2) the tails differ from tidal bridges, i.e. there is no optical counterpart at the tip of the 
tail down to $r\approx26$ mag~arcsec$^{-2}$ in the Sloan Digital Sky Survey (SDSS) images 
for extended features like the tails; 3) the projected length of the
tail is larger than the half of the symmetric part in H{\sc i} ($l_{t}>0.5D_{\rm HI}$;  
see Table 1).

Besides the similarities in gas morphology, the tails are also pointing roughly away from M87
as shown with arrows in Plate 1. These H{\sc i} tail galaxies are located at intermediate 
distances of $0.6-1$ Mpc from M87 in projection. All but NGC~4654 are found to the west of M87,
where the X-ray emission appears to be more elongated. They show a wide range of H{\sc i} gas
deficiency \citep[see][for the definition]{hg84}, from $-0.56$ to $0.8$. The H{\sc i} mass in
the tail varies from 10 to 40\% of the total. The optical and H{\sc i} properties are 
summarized in Table 1 and detailed descriptions of individual tails are given in the following 
section.

%\notetoeditor{Figure 1 should appear on the top-right side of pg. 2}
\begin{figure}
\epsscale{.94}
\plotone{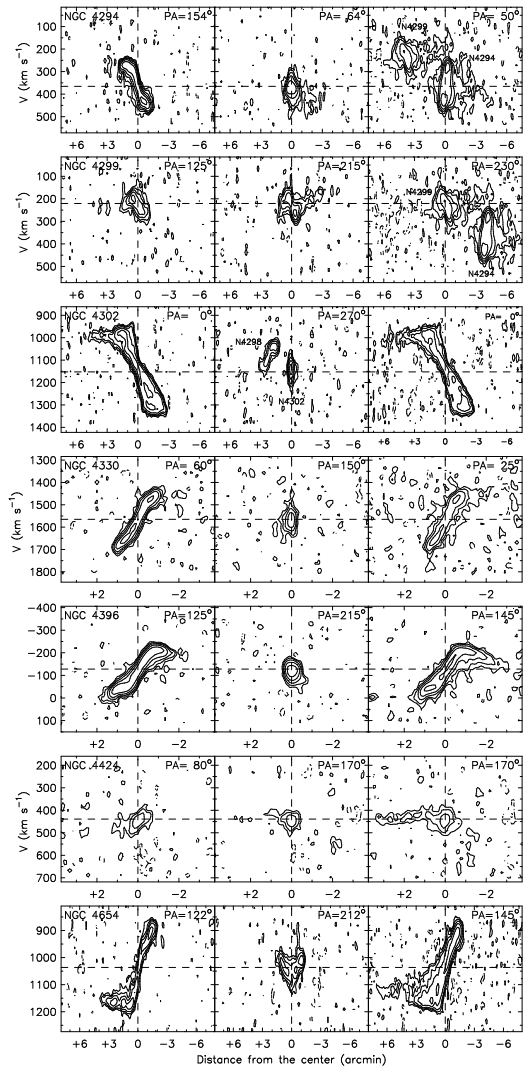}
\caption{For each galaxy, the first two figures from the left are
the PVDs on the disk along the major (left) and the minor (middle) axis. 
The figure on the right shows the PVD summed over the tail to the perpendicular direction of 
the tail. The angle of the cut is indicated in top-right corner. The dashed line represents
the center in the position and the velocity. From top to bottom, the widths of the cuts are
4.0$'$, 3.3$'$, 2.2$'$, 2.5$'$, 2.3$'$, 2.7$'$, and 3.9$'$, the contours for the PVDs along
the major/minor axis are -4, -2, 2, 4, 8, 16, 32..., $\times$0.25, 0.25, 0.26, 0.26, 0.24, 
0.29, 0.38~mJy~beam$^{-1}$, and for the PVDs along the tails, (same)$\times$4.0, 3.0, 2.5, 
2.5, 3.0, 4.0~mJy~beam$^{-1}$.\label{fig-tailxv}}
\end{figure}

\subsection{Comments on Individual H{\small I} Tail Galaxies}
\noindent{\bf NGC~4299/4294}:
These two galaxies are separated by 5.7$'$ ($\approx27$ kpc) in projection and 
$\approx120$~km~s$^{-1}$ in velocity. 
The tail of NGC~4294 is almost perpendicular to the stellar (and the H{\sc i}) disk and 
pointing to the SW. NGC~4299 has one tail pointing to the SW,  parallel to NGC~4294's tail and 
another tail pointing to SE. The second tail of NGC~4299 is much broader and lower in H{\sc i} 
surface density than the other one. The combined H{\sc i} mass of NGC~4299's tails is
$\gtrsim 4\times10^8~M_\odot$, $\approx36\%$ of the total emission. The H{\sc i} 
position-velocity diagram (PVD, Figure \ref{fig-tailxv}) shows a 2$\sigma$ connection between
the SE end of NGC~4294's disk, where the stellar disk is more extended. We see no optical 
counterpart along the H{\sc i} tails or between the two galaxies down to the limit mentioned. 

\smallskip
\noindent{\bf NGC~4302}:
The H{\sc i} is mildly truncated within the stellar disk in the south 
($R_{\rm HI}/R_{25}\approx1$, but a faint stellar envelope is seen beyond the H{\sc i} disk)
and the gas tail is extended to the north with no optical counterpart. A nearby galaxy, 
NGC~4298 is only 2.4$'$ ($\approx11$~kpc) to the west at a similar velocity 
($\Delta v\approx30$ km~s$^{-1}$). The H{\sc i} of NGC~4298 is more extended to the NW while
its stellar disk is more extended to the SE. The two galaxies overlap in velocity but there is
no H{\sc i} in between down to 2$\sigma$. While NGC~4298 looks disturbed both optically and in
H{\sc i}, NGC~4302 does not show any obvious signatures of a tidal disturbance.

\smallskip
\noindent{\bf NGC~4330}: 
Its H{\sc i} gas is  truncated within the stellar disk to the NE 
($R_{\rm HI}/R_{25}\approx0.75$) and the H{\sc i} tail extends to the SW. A deeper optical image
shows that also to the SW the undisturbed stellar disk extends well beyond the H{\sc i} tail 
(Abramson et al., in preparation). We do not see any obvious optical counterpart along the 
H{\sc i} tail but its GALEX (Galaxy Evolution Explorer) image shows a $UV$ tail from the SW 
end of the optical disk along the H{\sc i} to the NW, which is displaced from the optical disk.

\smallskip
\noindent{\bf NGC~4396}: 
The H{\sc i} morphology resembles that of NGC~4330, except that at the opposite side of the 
H{\sc i} tail (the SE) the H{\sc i} extends as far as the stellar disk. At that side the 
H{\sc i} is compressed and H$\alpha$ emission from star formation is enhanced. Along 
the tail no stellar light or $UV$ is seen down to the SDSS and GALEX limits.

\smallskip
\noindent{\bf NGC~4424}:
The H{\sc i} is severely truncated within the optical disk ($D_{\rm HI}/D_{25}\approx0.5$) and 
the tail extends from the south to SW. The tail shows a weak velocity gradient (Fig. 
\ref{fig-tailxv}) toward M49 as well as M87. The tail is clearly detected out to $\sim 10'$
($47$ kpc), although there is a hint of faint gas even further to the SW in the direction of
M49, which itself is the center of a subcluster of galaxies (Plate 1). The stellar morphology
and kinematics appear to be strongly disturbed by a gravitational interaction
\citep{kenney96,cortes06}. 

\smallskip
\noindent{\bf NGC~4654}: 
This is the only galaxy in the tail sample that also has a strong asymmetry in the stellar disk.
The H{\sc i} shows a compressed edge on one side and a long tenuous tail on the other side. SDSS
images show that starlight extends beyond the ridge of compressed H{\sc i} in the west,
implicating an ICM-ISM interaction as \citet{pm95} suggested.

\section{DISCUSSION}
\label{sec-discuss}
The tail galaxies are located at intermediate distances from the cluster center at a projected
distance of 0.6-1~Mpc from M87. The length and the width of the tails vary, but the tails are
all pointing roughly away from the cluster center (M87), suggesting that a single global
mechanism such as the cluster potential or ram-pressure due to the ICM might be responsible for 
these H{\sc i} tails. 

The cluster potential however, is not enough to affect galaxies significantly at those 
distances. The tidal acceleration due to the cluster potential ($\approx{2GMR}/{d^3}$, 
$d \gg R$ where $G$ is the gravitional constant, $M$ is the cluster mass within a radius $d$,
and $R$ is the size of the galaxy) is typically $\sim$2~(km~s$^{-1}$)$^2$~kpc$^{-1}$ while the 
gravitational acceleration of the galaxy ($V_{rot}^2/R$ where $V_{rot}$ is the galaxy's
rotational velocity) is always larger than the cluster potential by 3 orders of magnitude, with
a typical value of $5\times10^3$~(km~s$^{-1}$)$^2$~kpc$^{-1}$. 

%\notetoeditor{Figure 2 should appear on the right side of pg. 3, right below the table}
\begin{figure}
\epsscale{0.92}
\plotone{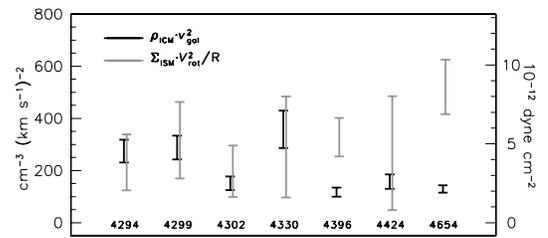}
\caption{The range of ram-pressure due to the ICM and the restoring force per unit area on the
disk for tail galaxies are presented with a bar. \label{fig-prange}}
\end{figure}
 
Alternatively galaxies can be gas stripped by ram-pressure when $\rho_{_{\rm ICM}} 
\cdot v_{gal}^2 > \Sigma_{_{\rm ISM}} \cdot V_{rot}^2/R$, where $\rho_{_{\rm ICM}}$ is the ICM 
density, $v_{gal}$ is the galaxy's velocity $w.r.t.$ the ICM, and $\Sigma_{_{\rm ISM}}$ is the 
ISM surface density \citep[][see also Voller 2001]{gg72}. In Figure~2 we compare the estimated 
ram-pressure strength with the restoring force per unit area in the disk of each galaxy. Rather 
than using the observed radial velocities, we use the simulations of \citet{vollmer01} to
estimate the range of relative velocities a galaxy can have at a given projected distance from
the cluster center (1250 to 1900~km~s$^{-1}$). The error bar in the ram-pressure indicates this
range. The main uncertainty in the restoring force is the surface density of the gas at the time
of stripping. We considered two extremes, the mean H{\sc i} surface density in the tail, and the
peak value measured in the tail. Figure~2 shows that ram-pressure could exceed the restoring
force for 5 out of 7 galaxies.

As seen in Plate 1, NGC~4330, NGC~4424 and NGC~4302 (and its neigbour NGC~4298) clearly show
that their H{\sc i} disk is truncated to well within the stellar disk indicating that these 
three galaxies must have been affected by a gas-gas interaction. The systemic velocity of 
NGC~4330 and NGC~4302 is close to the cluster mean, indicating that they move mostly in the
plane of the sky. This combined with the fact that their H{\sc i} tails point away from the 
cluster center, suggests these galaxies are falling into the cluster along highly radial orbits.
NGC~4424, whose systemic velocity is highly blueshifted with respect to the cluster mean, could
be falling in from the back. Another possibility is that it has passed close to M49 and has been
stripped by the ICM of that subcluster. Its tail points directly toward the subcluster.
In this scenario the NGC~4424 tail would be more comparable to another tail in Virgo recently
found by \citet{ovg05}, which appears to be gas stripped from NGC~4388 by the hot ICM gas in 
the subgroup of M86. Similarly a passage close to M49 could have caused the gravitational 
disturbance of NGC~4424, as M86 might have gravitationally disturbed NGC~4388.

At the other extreme, the two galaxies for which our simple estimate of ram-pressure appears too 
low to have an effect, still show signs of gas-gas interaction. Both galaxies, NGC~4396 and NGC~4654, 
have H{\sc i} compressed on one side, while the H{\sc i} tail at the opposite side does not seem 
to have an optical counterpart. In deeper optical images, NGC~4654 reveals more stellar light 
beyond the H{\sc i} disk to the NW (\S~3.2). In fact \citet{vollmer03} concludes that the 
H{\sc i} morphology and kinematics of NGC~4654 can best be explained by the combined effect 
of ram-pressure stripping and a tidal disturbance by its companion, NGC~4639. On the other hand, 
NGC~4396 looks optically undisturbed. Several explanations are possible, either the galaxy is 
affected by viscous turbulent stripping instead of the simple momentum transfer, or the ICM 
pressure is enhanced due to bulk motions or local clumping of the gas, as it is likely in 
NGC~4522 (Kenney et al. 2004). 

Additionally, the kinematics of the tails also support an ICM-ISM interaction for most cases.
For all galaxies except NGC~4294 and possibly NGC~4330, the velocity gradient in the H{\sc i} 
tails is toward the cluster mean (Fig. \ref{fig-tailxv}) as it is expected if the H{\sc i} gas
has been stripped by the cluster gas.

We have seen for NGC~4654 that the tidal interaction with a nearby companion could have made 
the galaxy more vulnerable to ram-pressure stripping. We may well 
ask whether tidal interactions with neighbors could help explain the existence of the other tails.
\citet[][2004]{mihos01} shows that if galaxies fall in as groups, the effect of the cluster 
potential gets greatly enhanced. In Table 1 we list the nearest companion for each of the
galaxies. Almost all have neighbors that are close in projection ($<$100~kpc) and in velocity
($<$100~km~s$^{-1}$). In a dense cluster environment being close in projection does not 
necessarily imply physical association, on the other hand the cumulative effect of many distant
encounters can also loosen the outer parts (harassment). It is thus not unlikely
that gravity plays some role, yet morphologically all the systems show some 
evidence for gas-gas interactions as well. The only exception is the NGC~4299/4294 pair, 
although a parallel orientation of two tidal tails is rarely seen in interacting systems,
and the tails have no stellar counterparts down to the SDSS limit. In analogy with the 
\citet{mihos04} result tidal interactions may  make it more easy to ram-pressure strip galaxies 
by bringing material to the outer disk \citep{vh06}.

To conclude, while previous H{\sc i} imaging showed that highly H{\sc i} deficient galaxies 
in the center of Virgo have very small H{\sc i} disks \citep[e.g.][]{cayatte90}, we are now 
providing H{\sc i} images of galaxies that are beyond the strongly H{\sc i} deficient zone
\citep[$>0.5~$Mpc;][see also Fig.~3 in Solanes et al. 2001]{cayatte90}. Our images show how 
galaxies just outside the virial radius may begin to lose their gas. In this region (0.6$-$1~Mpc
from M87), seven of the 16 galaxies imaged by us (out of 27 spiral galaxies in the VCC catalog
with m$_p<$13.75, Binggeli, Sandage \& Tammann 1985) show H{\sc i} tails. Thus at least 26\%
of the large spiral galaxies in this region of Virgo seem to be recent arrivals being stripped
of gas. Interestingly, since it 
only affects the loosely bound gas in the outer parts of the galaxies, the simple momentum 
transfer ram-pressure picture still works for most of these galaxies, though viscous turbulent 
stripping, which can remove the gas more efficiently from edge-on encounters {$w.r.t.$} the 
ICM \citep{nulsen82}, could affect some galaxies (e.g. NGC~4302). It seems then that galaxy 
pre-processing occurs through mild gas-gas and/or tidal interactions before the galaxies enter
the dense part of the cluster. Once they are within a virial radius gas gets removed within a 
gigayear through momentum transfer ram-pressure stripping of the more higly radially infalling
galaxies and possibly viscous turbulent stripping of galaxies on more circular orbits.

\acknowledgments
This work has been supported in part by NSF grants to Columbia University and Yale University.

\clearpage
%\notetoeditor{Figure 3 (Plate 1) should appear at the end of the paper on the whole page as a plate.}
\begin{figure}
\epsscale{1.0}
\plotone{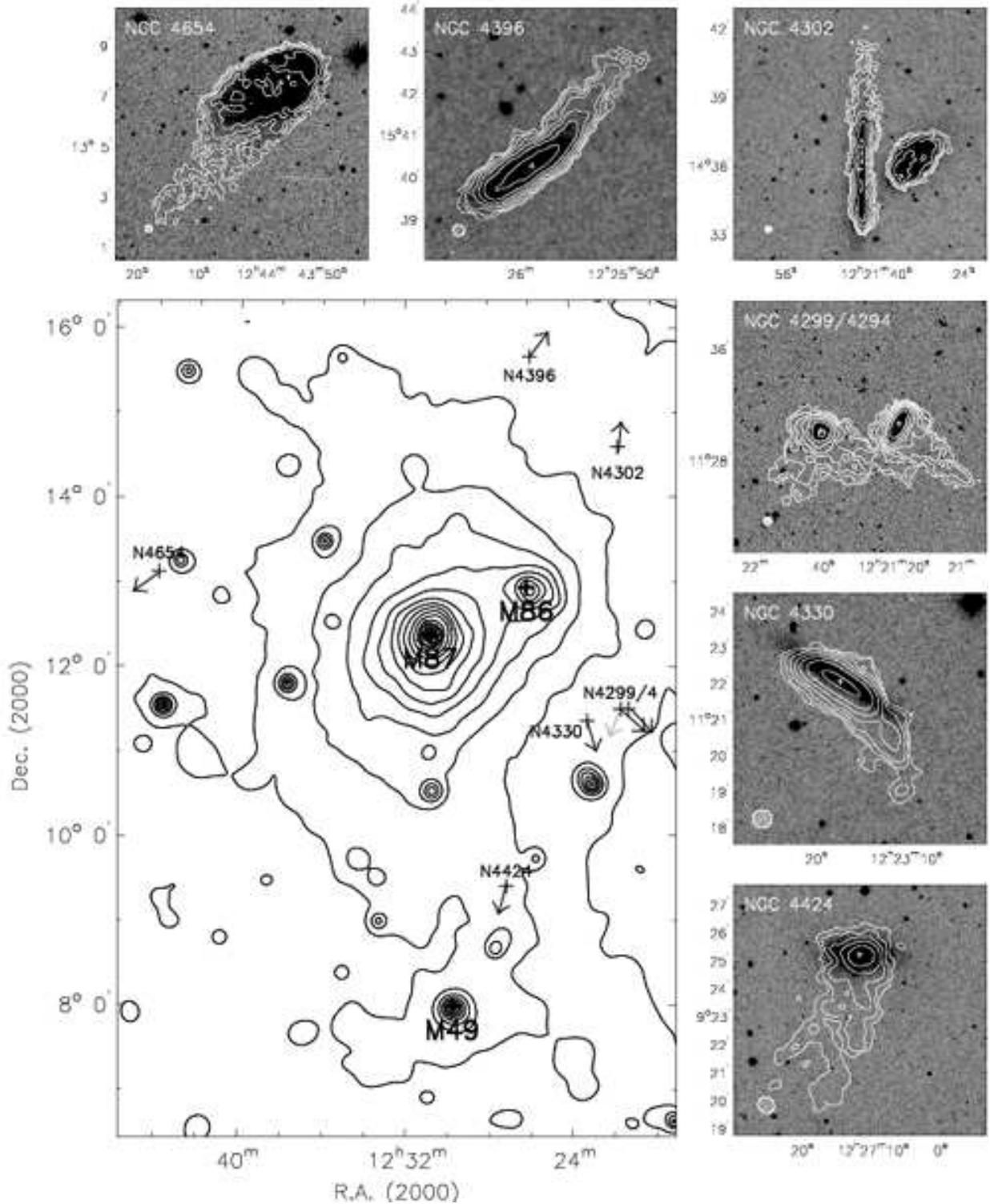}
\caption{(Plate~1) bottom-left) The locations of the H{\sc i} tail galaxies are shown with the 
cross on the X-ray background of the Virgo region \citep[0.5$-2.0$ keV, ROSAT;][]{bohringer94}.
The directions of the tails are indicated with the arrow. The second tail of NGC~4299 (E tail) 
is shown in lightgray. Seven figures on the top and on the right, we show zoomed views of 
individual galaxies. The H{\sc i} contours (white) are shown overlaid on the Digitized Sky
Survey (DSS) image in grayscale. The galaxy name and the synthesized beam size appear in the
upper-left and the bottom-left corner in each box. 
The white crosses indicate the optical center. The H{\sc i} contours are 2.8 (NGC~4294/9), 6.7 
(NGC~4302), 2.2 (NGC~4330), 4.3 (NGC~4396), 1.9 (NGC~4424), 13.0 (NGC4654)$\times$1, 2, 4, 8, 
16, ... in $10^{19}$~cm$^{-2}$. \label{fig-tail}}
\end{figure}

\end{document}